\documentclass[dvips,a4paper,12pt]{article}
\usepackage[dvips]{graphicx}

\begin{document}
\pagestyle{empty}

\begin{center}

{\bf \large Swelling and dissolution of onion phases:\\ The effect of
temperature} 

\vspace{1cm}
{\bf M. Buchanan$^\dagger$, S.U. Egelhaaf and M. E. Cates}

\vspace{1cm}
Department of Physics and Astronomy, The University of Edinburgh, \\ Mayfield
Road, Edinburgh, UK EH9 3JZ\\
\vspace{1cm}
$\dagger$ Present address: Centre de Recherche Paul-Pascal, CNRS, \\ Avenue du
Docteur-Schweitzer, F-33600 Pessac, France\\
\vspace{1cm}
\today
\end{center}

\vspace{2cm}

\begin{abstract}

Contact experiments have been performed between an onion lamellar phase and
brine, in the SDS/octanol/brine system.  Using video microscopy we have studied
the nonequilibrium behaviour of the swelling and dissolution process of onions.
Experiments at $\textrm{T} = 20^{\circ}\textrm{C}$ and $30^{\circ}\textrm{C}$
showed that temperature has a strong effect on their behaviour. At low
temperature onions are observed to diffuse away from the onion phase and only
swell sightly. However by increasing the temperature we induce the formation of
sponge phase ($L_3$) at the onion/brine interface. Onions that initially swell
then dissolve into $L_3$~phase expel a stable core which moves to the micellar
phase and remains there. Over a longer period of time (several days) we have
also observed coalescence leading to the formation of large onions of up to
$\sim 100 \mu \textrm{m}$ diameter.  These huge onions have a radial
distribution of domains, or solvent cavities, within them.

\end{abstract}

\newpage
\pagestyle{plain}
\setcounter{page}{1}

\section{Introduction}

Surfactant phases can exhibit a variety of interesting behaviour when observed
in nonequilibrium experiments \cite{egelhaaf:kinetics}. Specifically, we are
interested in the relaxation behaviour of onion phases. Onions consist of
several closed bilayer shells, i.e. multilayer
vesicles~\cite{diat:shearrate}. Their equilibrium properties have been
extensively studied with different techniques as well as their flow behaviour
using
rheology~\cite{diat:shearrate,hoffmann:shearinf,hoffmann:vesicle,hoffmann:charge,hoffmann:surfactant,
leng:necks,panizza:onions}.  However, only little is known on their behaviour
during relaxation to equilibrium. In particular, we investigate the swelling
and dissolution of onion phases after concentration quenches. Depending on the
conditions onions are expected to swell and to dissolve eventually forming a
micellar ($L_1$) and/or sponge phase ($L_3$).

Penetration-scan experiments \cite[pp.~535--538]{laughlin:book} provide a
useful method for studying surfactant phase behaviour and kinetics
\cite{miller:review,miller:diffusion,miller:solubilization,mori:dynamics,hakemi:diffusion}.
These experiments involve placing one phase, e.g. $L_{\alpha}$, in a capillary
tube and contacting it with another phase, e.g. solvent such as water or
brine. At the interface between the two phases a sequence of intermediate,
metastable phases may then arise. The structural arrangement of the mesophases
is then observed using polarisation microscopy, which also facilitates to
identified the formed mesophases by the birefringent
textures~\cite{kleman:points}.  This technique has recently been used to study
the formation and swelling of a lamellar phase by contacting pure surfactant
with an aqueous phase
\cite{argen:DoBSphase,argen:lamellar,Buchanan:myelins}. In many contact
experiments, a crystal or neat surfactant liquid phase is used instead of the
lamellar phase. However, in most of these cases a lamellar phase forms almost
immediately at the interface, and controls the subsequent evolution. These
random oriented bilayers (polycrystalline powder lamellar phase) are not very
well-defined microstructures. We thus prefer to perform contact experiments
directly with well-characterised lamellar samples, e.g consisting of onions
with a certain size, than with pure surfactant from which the lamellar phase
forms in a rather uncontrollable fashion. We exploit the fact that by shearing
a lamellar phase its microstructure is changed from the powder state to a close
packed array of onions~\cite{diat:shearrate}.

In this study we perform penetration scans with an onion phase at different
temperatures. Changing the temperature of a bilayer system affects various
properties of the bilayers, such as fluctuations, bilayer topology and density
of defects. This determines the equilibrium phase behaviour and also has
important consequences for the non-equilibrium dynamics. In this paper we
investigate non-equilibrium behaviour, specifically the swelling and
dissolution of onions upon addition of brine.  At low temperature onions are
observed to disperse and swell slightly when observed over $4$ days. Over the
same period of time, but at a higher temperature, a variety of phenomenon can
been observed. Initially, the onions disperse and then proceed to swell up to
three times their original diameter. The centre of the onion has a core of much
larger density than the rest of the onion. Eventually, the onion dissolves
forming a sponge phase which can be observed in abundance at the onion/micellar
interface. During onion dissolution the core is expelled which then proceeds to
diffuse away from the sponge phase until it eventually dissolves in the
micellar phase. After $4$ days some onions are observed to coalesce forming
large onions with diameters of about $100 \mu \textrm{m}$.

\section{Experimental Section}
\label{expsec}

	\subsection{Materials}					\label{exp:mat}

SDS was supplied by Touzard and Matignon, France, ($99\%$ purity) and used
without further purification.  The lamellar phase was prepared by dissolving
$9\%$ SDS and $11\%$ octanol in brine ($20$g/l NaCl in distilled water).
This was then left to equilibrate, until the lamellar phase was homogeneous, 
which took approximately two weeks.

	\subsection{Preparation of onion phase}

To prepare the onion phase, the lamellar phase is sheared in a temperature
controlled transparent Couette (light-scattering) cell with a $1$mm gap,
manufactured by Caplim.  In order to obtain onions that are $10\mu \textrm{m}$
in diameter the lamellar phase is initially loaded into the cell and then
heated to $\textrm{T} = 35^{\circ}\textrm{C}$, where the lamellar phase is
known to undergo a transition to the sponge phase.  The sample is then
subjected to a large shear rate of about $200 \textrm{s}^{-1}$) while the
temperature is lowered to $\textrm{T} = 30^{\circ}\textrm{C}$.  When onions
form the light scattered from the sample gives a characteristic ring in the
forward direction whose radius is related to the onion
size~\cite{diat:shearrate,bergenholtz:onions}. A steady state is assumed to
have been reached when no further change in the size of the light-scattering
ring is noticed. The sample is then removed from the Couette cell and loaded
into the capillary tube or glass slide and cover slip arrangement as described
below.

	\subsection{The contact experiment}		\label{contexp}

The phase penetration scan experiment involves contacting onion phase with
brine. This was either done using a glass slide and cover slip or a capillary
tube.  In the first case the onion phase is placed between a glass slide and a
cover slip, which is raised at one edge by a spacer, and then contacted with
brine. This method has the advantage that it allows to produce thin samples and
therefore the contrast is high and structures can be studied in detail. In the
second method, the capillary tube ($4$mm $\times$ $0.4$mm) is gently inserted
into the onion phase whereupon capillary action loads the sample into the
tube. At the sample end the tube is then sealed, using Araldite glue.
Subsequently the sample is observed under a microscope to check that loading
did not introduce disruption of the microstructure in the onion phase.  The
sample is then contacted with brine. This is done outside the microscope using
a fine pipette in a controlled and gentle fashion. The sample is then placed on
a microscope slide and slightly tilted at one end to keep any trapped air
bubbles away from the lamellar/water interface. The capillary tube is
completely sealed to prevent any evaporation. Thus this allows us to study the
sample for up to 4 days. Disruption of the sample often occurs, when some of
the material becomes torn from the interface as water flows in to displace the
air. These samples are rejected and only those are further studied which
produce a smooth initial interface between onion and aqueous phases.

	\subsection{Microscopy}

Observations were made using an Olympus BX50 microscope in bright-field
mode between crossed polarisers with long working distance objectives. In 
a few cases we applied Normarski DIC which detects changes in
refractive index in the sample and gives better contrast in experiments
where we use glass slide and coverslip (section~\ref{contexp}).  
The evolution of the samples is recorded using a CCD camera and
time lapse video recorder together with a framegrabber PC. The contrast of the
presented images has been enhanced by thresholding. All experiments were done
at a well-defined temperature in a Linkam LTS hot-stage which ensured
controlled and stable conditions.\\
\\

%\newpage

\section{Results}
\label{results}

Samples composed of onions are contacted with brine at two temperatures,
$\textrm{T} = 20^{\circ}\textrm{C}$ and $30^{\circ}\textrm{C}$, to study
temperature effects on swelling and dissolution of lamellar phase.

\subsection{Low temperature $\textrm{T}=20^{\circ}\textrm{C}$}	\label{sec:lowT}

Onions with a characteristic diameter of $10 \mu \textrm{m}$ were contacted
with brine.  The behaviour of the onions following this quench was studied
using two different geometries. The first method is optimised for a high
resolution of the onion organisation, which is achieved preparing thin
samples. The sample is placed between a glass slide and coverslip and then
observed using Normarski DIC. Upon contact with brine, onions in the bulk far
from the interface remain close packed.  Figure~\ref{dic}A shows the hexagonal
organisation of the onion phase immediately after preparation.  They are
spontaneously deformed into this shape due to the low curvature energy cost of
the bilayers. After a short time, about 2 minutes, the onions come apart and
swell while becoming spherical (figure~\ref{dic}B).

The same experiment was also performed in a capillary tube. Because the
capillary tube can be sealed, this allows us to study the long time behaviour
without being affected by evaporation. We followed the behaviour of onions
close to the interface, but still in bulk. Four days after contact these onions
have moved apart, while they are not yet free from the bulk
(figure~\ref{dic}C).  Outside each onion we have observed a `halo' domain. The
walls of these domains are in contact with each other and are observed directly
between the onions.

Directly at the interface the onions initially cascade from the interface. Due
to their lower density, they move onto the top surface of the glass capillary
tube (figure~\ref{lowtemp}A).  As the onions detach from the interface, they
swell slightly from initially about $10 \mu \textrm{m}$ to approximately $12-14
\mu \textrm{m}$ (figure~\ref{lowtemp}B).  The onions are not found to coalesce
and appear to be fairly monodisperse across the whole interface. Furthermore,
the internal structure of the onions appears to remain homogeneous throughout
the observation period.

%\newpage
\subsection{High temperature $\textrm{T} = 30^{\circ}\textrm{C}$}		\label{sec:highT}

We have also performed penetration scan experiments on onion samples at a
higher temperature $\textrm{T} = 30^{\circ}\textrm{C}$. After contact with
brine, the onions detach from the onion/brine interface and move to the top of
the glass capillary tube (figure~\ref{hightemp1}A). In the region of detached
and swelling onions, sponge phase ($L_3$) nucleates, at first forming droplets
on the capillary tube surface.  The $L_3$~droplets then coalesce to form a
large region of $L_3$ phase (figure~\ref{hightemp1}B).  Since the SDS/octanol
bilayers have a lower density than brine we observe the $L_3$~phase on the top
surface of the capillary tube.

During the dissolution of onions into the $L_3$~phase, domains within the onion
are observed (figure~\ref{ring}). An inner and outer part of the onion can be
distinguished as well as a core. The onion core albeit peculiar seems to be no
artifact, because it has been observed in various other
systems~\cite{oda:cores,buchanan:AOT}.  As the onion dissolves the interface
between the inner and outer domains moves towards the centre of the
onion. After the onion dissolves all that is left is a small core of
approximately $1\mu \textrm{m}$ size. The core undergoes Brownian motion and eventually
leaves the $L_3$~phase. Upon leaving the $L_3$~phase the core continues to
undergo more vibrant Brownian motion in the micellar phase until it dissolves
about $10$s later.  Figure~\ref{schemsds} shows a schematic representation of
the vertical profile in the capillary tube.

After $1$ day we observe coalescence, where the onions merge together to form
larger onions. Since this process is slow, onions can be observed at different
stages of coalescence (figure~\ref{hightemp2}).  This behaviour is very
different from the previous observations at $\textrm{T} =
30^{\circ}\textrm{C}$. In figure~\ref{hightemp2}C-E we observe one of the
largest onions in the system 4 days after contact.  At this point in time we
only observe a few very large onions since most of the small onions have fused
with these.

After coalescence, onions are often observed to show several concentric domains
with sharp interfaces.  We also observe solvent-like-domains at some of these
interfaces.  The centre of the onion in this case is very brightly
birefringent. This may imply a dense lamellar phase at the core. On close
inspection a small inner core in the centre of a bright birefringent lamellar
shell is observed (see figures~\ref{hightemp2}C-E).  This core is mobile in the
centre of the onion and can be observed to undergo Brownian motion while being
confined by the shell.

\subsection{Dilution experiments}

In a separate experiment, onions of $10 \mu \textrm{m}$ diameter are added to
solvent and then dispersed by gently shaking the sample. The diluted onion
sample is placed in a flat capillary tube and observed at $\textrm{T} =
30^{\circ}\textrm{C}$ for a period of 6 hours.  The onions appear to be
slightly swollen; some of the onions are as big as 18$\mu \textrm{m}$. With
these onions we observe swirly defects within the onions
(figure~\ref{quen}A). If the onions are kept at high temperature ($\textrm{T} =
30^{\circ}\textrm{C}$) the onions dissolve after about one day in the brine
without further swelling.  However, if the temperature is lowered to
$\textrm{T} = 20^{\circ}\textrm{C}$, the onions not only not dissolve, but in
addition the defects vanish over time (figure~\ref{quen}B-D). We also noted a
slight, but consistent decrease in onion size during this process.

%-------------------------------------------------------------------------%
\section{Discussion}				\label{sdsdisc}
%-------------------------------------------------------------------------%

        \subsection{Onion swelling}

An onion phase was contacted with brine at two different temperatures. The
swelling and dissolution behaviour was found to depend on temperature and
dramatic changes were found. When observed over a one week period at low
temperature ($\textrm{T} = 20^{\circ}\textrm{C}$) onions remain intact while
they detach and diffuse away from the interface. They do not appear to swell by
a large amount. In contrast, onions at high temperature ($\textrm{T} =
30^{\circ}\textrm{C}$) can swell up to three times their diameter and
eventually `melt' into $L_3$~phase. This can be rationalised based on the
equilibrium phase behaviour. A lamellar phase can only swell, i.e.  accommodate
solvent, to a maximum bilayer spacing which is determined by the position of
the $L_{\alpha}$~phase boundary. When it reaches this point the bilayers will
reorganise their structure and form sponge phase. At high temperature the
maximum swollen lamellar phase can reach a larger layer spacing before
transforming to $L_3$~phase than at lower temperatures. This behaviour is
reflected in the observed swelling of the onions at higher temperature.

        \subsection{Onion collapse and cores} 	\label{sdscores}

At low temperature onions only swell slightly and cores are not
observed. However, in the high temperature experiments onions are
found to eventually dissolve into $L_3$~phase and expel a stable
core. While stable cores have been observed in
nonionic/water~\cite{oda:cores} and AOT/brine~\cite{buchanan:AOT}
mixtures, the present system has some unique features that have not
been observed before. During the onion collapse a domain is observed
in the onion which shrinks towards the core (figure~\ref{ring}).  The
core will remain stable only for several minutes and then dissolve in
the $L_1$~phase. There are two possible theoretical explanations for the
stability of the core.

First we consider a possible kinetic reason for core stability.  In order for
onions to swell, solvent has to be transported through the layers into the
onion. This requires defects, usually necks~\cite{roux:sponge}, which nucleate
between layers. We can try to estimate the number of defects as a function of
the diameter of one shell in the onion, i.e. one bilayer. The density of
defects between bilayers was measured using rheology and found to increase
drastically with increasing temperature~\cite{leng:necks}.  In particular, at
low temperature ($\textrm{T} = 20^{\circ}\textrm{C}$) the area density of necks
is small ($\phi=0.02\%$), whereas at a higher temperature ($\textrm{T} =
30^{\circ}\textrm{C}$) it is significantly larger ($\phi=0.4\%$).  The diameter
of the neck is expected to be of order of the layer thickness, $d \sim 5 \times
10^{-3} \mu \textrm{m}$, which gives the defect area to be $A_d \sim 2 \times
10^{-5} \mu \textrm{m}$. The number of defects $n$ on an onion layer with
diameter $D$ is thus estimated as $n \sim D^2 \phi / A_d$.  Using the
experimentally measured values of $\phi$ at $\textrm{T} = 30^{\circ}\textrm{C}$
it follows that for onions smaller than $\sim 0.1 \mu \textrm{m}$ there is on
average less than one defect and thus swelling might not occur or be very slow.
That this size is smaller than the core diameters observed, possibly indicates
that more than one defect is required to swell the onion within a reasonable
time.

Alternatively for swelling to occur, the distance between bilayers has to
increase, while still maintaining the layered structure. To ensure that the
layers are not in a bound state, but wish to swell thermodynamically, thermal
undulations must be present. In contrast to the lamellar phase, the transition
from a bound to an unbound state within the onion has to allow for a tension
across the membrane~\cite{diamant:cores}.  At the core, high curvature can
induce a tension which suppresses membrane undulations and thus leads to a
bound state.  Surfactants in membranes with high curvature, at the core, will
diffuse to membranes with a lower curvature. The tension of membranes in the
centre will hence be higher than in the rest of the onion.  This idea has been
developed quantitatively by Diamant~\cite{diamant:cores} and successfully
predicts that the core will remain in a bound state during the swelling of the
onion. Figure~\ref{haimscore} shows the calculated density profile of a swollen
onion where the bright region corresponds to high density and the dark region
to low density.  This argument assumes local equilibration of the bilayers
within the onion during the swelling process and is thus, in contrast to the
first argument, based on thermodynamic, equilibrium principles. Later we give
evidence which supports this point of view (section~\ref{disc-dil}).

        \subsection{Onion coalescence}

Giant onions have been observed as large as $100 \mu \textrm{m}$. They can be
seen in penetration scans 3-4 days after contact and are the result of onions
coalescing (figure~\ref{hightemp2}). In emulsions and vesicles it is well known
that droplets can coalesce and merge to form bigger aggregates. The time to
coalesce is dependent on the {\it energy barrier} that must be surmounted in
order to join the surfactant layers, which might either be a mono- or bilayer
in the case of emulsions and vesicles, respectively. Once over the barrier they
then rapidly merge~\cite{bibette:emulsions,burger:membrane}. In the case of
(multilamellar) onions one layer has to connect at a time. For the outer onion
layer to connect, a neck must form between the two outer bilayers. Once this
barrier is passed the onion cannot relax quickly since the next layer has to
form a neck and so on. The complete process of coalescence has been observed
(figure~\ref{hightemp2}) and takes several days to complete.  Also domains or
cracks are observed radially in these onions.  These may be remains of onions
that have joined. After they have coalesced fully they leave a scar behind
where there is some density variation in the onion. Alternatively there could
be thermodynamic reasons for a radial domain structure in some
cases~\cite{diamant:cores}.

        \subsection{Onion dilution}		\label{disc-dil}

After the dilution of onions with brine, we have observed defects in the
onions.  These defects, which are probably solvent domains have a
characteristic `swirl'. They may be due to the osmotic shock after the onions
have been dispersed. Over 4 days these defects disappear, if the onions are
kept at a low temperature of $20^{\circ}\textrm{C}$. (In contrast, at
$30^{\circ}\textrm{C}$ the onions dissolve before the defects can disappear.)
Since the onions slightly shrink this suggests that solvent is expelled from
the onion. The bilayers within the onion are thus able to reorganise into an
energetically more favourable state where there are no internal defects and
thus the bilayers can reach a local equilibrium state. This reorganisation does
not seem to depend on neck-like defects, since the density of these defects is
small at low temperature~\cite{leng:necks}. This also has important
consequences for the stability of cores discussed above
(section~\ref{sdscores}).  This observation makes the kinetic reason involving
necks less likely and thus supports the thermodynamic explanation based on the
unbinding theory of an onion.

\section*{Acknowledgements}

We are grateful to J. Leng, H. Diamant, P. Garrett, D. Roux, J. Walsh and
P. Warren for useful discussions.  One of us (M. B.) thanks Unilever PLC and
EPSRC (UK) for a CASE award.  Work funded in part under EPSRC Grant GR/K
59606.

%\bibliography{../../bibliography/general}

\begin{thebibliography}{10}

\bibitem{egelhaaf:kinetics}
S.U. Egelhaaf.
\newblock Kinetics of structural transitions in surfactant solutions.
\newblock {\em Current opinion in Colloid and Interface Science},
  3(6):608--613, 1998.

\bibitem{diat:shearrate}
O.~Diat, D.~Roux, and F.~Nallet.
\newblock Effect of shear on a lyotropic lamellar phase.
\newblock {\em J. Phys. II France}, 3(9):1427 -- 1452, 1993.

\bibitem{hoffmann:shearinf}
M.~Bergmeier, M.~Gradzielski, H.~Hoffmann, and K.~Mortensen.
\newblock Behavior of ionically charged lamellar systems under the influence of
  a shear field.
\newblock {\em Journ. Phys. Chem.}, 103(9):1605--1617, 1999.

\bibitem{hoffmann:vesicle}
H.~Hoffmann and W~Ulbricht.
\newblock Vesicle phases of surfactants and their behaviour in shear flow.
\newblock {\em Tenside Surfactants Detergents}, 35(6):421--438, 1998.

\bibitem{hoffmann:charge}
M.~Bergmeier, M.~Gradzielski, H.~Hoffmann, and K.~Mortensen.
\newblock Behavior of a charged vesicle system under the influence of a shear
  gradient: A microstructural study.
\newblock {\em Journ. Phys. Chem. B}, 102(16):2837--2840, 1998.

\bibitem{hoffmann:surfactant}
H.~Hoffmann, C.~Thunig, P.~Schmiedel, and U~Munkert.
\newblock Surfactant systems with charged multilamellar vesicles and their
  rheological properties.
\newblock {\em Langmuir}, 10(11):3972--3981, 1994.

\bibitem{leng:necks}
J.~Leng, P.~Sierro, F.~Nallet, and D.~Roux.
\newblock Defects-density measurement in a lyotropic lamellar phase using onion
  texture.
\newblock to be published.

\bibitem{panizza:onions}
P.~Panizza, D.~Roux, V.~Vuillaume, C-Y.D. Lu, and M.E. Cates.
\newblock Viscoelasticity of the onion phase.
\newblock {\em Langmuir}, 12(2):248 -- 252, 1996.

\bibitem{laughlin:book}
R.G. Laughlin.
\newblock {\em The Aqueous Phase Behaviour of Surfactants.}
\newblock Academic Press, 1994.

\bibitem{miller:review}
C.A. Miller.
\newblock Spontaneous emulsification produced by diffusion - {A} review.
\newblock {\em Colloids and Surfaces}, 29:89 -- 102, 1988.

\bibitem{miller:diffusion}
C.A. Miller and K.H. Raney.
\newblock Diffusion path analysis of dynamic behaviour of oil-water-surfactant
  systems.
\newblock {\em AIChE Journal}, 33:1791--1799, 1987.

\bibitem{miller:solubilization}
C.A. Miller and K.H. Raney.
\newblock Solubilization emulsification mechanism of detergency.
\newblock {\em Colloids and {Surfaces} {A}-{Physicochemical} and {E}ngineering
  {Aspects}}, 74(2):169, 1993.

\bibitem{mori:dynamics}
F.~Mori, J.C. Lim, O.G. Raney, C.M. Elsik, and C.A. Miller.
\newblock Phase-behaviour, dynamic contacting and detergency in systems
  containing triolien and nonionic surfactants.
\newblock {\em Colloids and {Surfaces} {A}-{Physicochemical} and {E}ngineering
  {Aspects}}, 40(3):323 -- 345, 1989.

\bibitem{hakemi:diffusion}
H.~Hakemi, P.P. Varanasi, and N.~Tcheurekdjian.
\newblock Diffusion studies in liquid crystalline phase of surfactant
  solutions.
\newblock {\em J. Phys. Chem.}, 91(1):120--125, 1987.

\bibitem{kleman:points}
M.~Kleman.
\newblock {\em Points, Lines and Walls}.
\newblock Wiley-Interscience Publication, 1983.

\bibitem{argen:DoBSphase}
A.~Sein and J.B.F.N. Engberts.
\newblock Lyotropic phases of {D}o{B}{S} with different counterions in water.
\newblock {\em Langmuir}, 12(12):2913--2923, 1996.

\bibitem{argen:lamellar}
A.~Sein and J.B.F.N. Engberts.
\newblock Formation of a lamellar phase : Rearrangement of amphiphiles from the
  bulk isotropic phase into a lamellar fashion.
\newblock {\em Langmuir}, 12(12):2924--2931, 1996.

\bibitem{Buchanan:myelins}
M.~Buchanan, S.~Egelhaaf, and M.E. Cates.
\newblock Dynamics of interface instabilities in a nonionic lamellar phase.
\newblock to appear in Langmuir.

\bibitem{bergenholtz:onions}
J.~Bergenholtz and N.J. Wagner.
\newblock Formation of {A}{O}{T}/{Brine} {Multilamellar} {Vesicles}.
\newblock {\em Langmuir}, 12(13):3122 -- 3126, 1996.

\bibitem{oda:cores}
R.~Oda.
\newblock {\em X-ray Diffraction Study of a Three-Component Lamellar Phase}.
\newblock PhD thesis, University of Tokyo, 1988.

\bibitem{buchanan:AOT}
M.~Buchanan, J.~Arrault, and M.E. Cates.
\newblock Swelling and dissolution of lamellar phases: The role of bilayer
  organization.
\newblock {\em Langmuir}, 14(26):7371 -- 7377, 1998.

\bibitem{roux:sponge}
D.~Roux, C.~Coulon, and M.E. Cates.
\newblock Sponge phases in surfactant solutions.
\newblock {\em The Journal of Physical Chemistry}, 96(11):4174--4187, 1992.

\bibitem{diamant:cores}
H.~Diamant.
\newblock {\em Problems in Self-Assembly of Amphiphilic Molecules and
  Polymers}.
\newblock PhD thesis, Tel-Aviv University, 1999.

\bibitem{bibette:emulsions}
J.~Bibette, Leal~F. Calderon, and P.~Poulin.
\newblock Emulsions: basic principles.
\newblock {\em Rep. Prog. Phys.}, 62:969 -- 1033, 1999.

\bibitem{burger:membrane}
Koert~N.J. Burger.
\newblock {\em Morphology of membrane fusion}, volume~44 of {\em Current topics
  in membranes}, chapter~11.
\newblock Academic Press, 1997.

\end{thebibliography}

%\newpage
\begin{figure}[!h]
\centering
\caption[Onion phase contacted with brine and observed using Normarski DIC.]{
Onions in the bulk of the sample after contact of the onion phase with brine at
$\textrm{T}=20^{\circ}\textrm{C}$.  (A) Hexagonal close packed order before
contact.  (B) About two minutes after a small dilution, onions are
observed to swell slightly.  (C) Four days after contact `halo' domains
around the onions are observed in the bulk close to the interface
(arrows). Images (A) and (B) are obtained using a glass slide and coverslip and
observing with Normarski DIC, while image (C) results from an experiment with a
capillary tube using brightfield microscopy.}
\label{dic}
\end{figure}

%\newpage
\begin{figure}[!hbt]
\centering
\caption{Onions at the onion/brine interface at
$\textrm{T}=20^{\circ}\textrm{C}$. 
(A) After several hours onions detach from the interface. 
(B) Within $4$ days onions swell by a small amount.}
\label{lowtemp}
\end{figure}

%\newpage
\begin{figure}[!hb]
\centering
\caption{ Formation of sponge phase ($L_3$) after onion phase (right) is 
contacted with brine (left) at $\textrm{T} = 30^{\circ}\textrm{C}$.  
(A) Interface immediately after contact.  (B) Sponge phase begins to form at
the onion/brine interface 3 hours after contact.}
\label{hightemp1}
\end{figure}

%\newpage
\begin{figure}[!h]
\centering
\caption{ Optical micrograph of onion kinetics at the micellar/sponge
($L_1$/$L_3$) interface.  Onions (left) dissolve into a region of sponge phase 
(right). Arrow indicates the progress of a particular onion 
over time (times shown in minutes).}
\label{ring}
\end{figure}

%\newpage
\begin{figure}[!h]
\centering
\caption{Schematic diagram of an onion-brine contact experiment at $\textrm{T} = 30^{\circ}\textrm{C}$.  
Vertical cross section of the capillary tube after the $L_3$~phase has formed.}
\label{schemsds}
\end{figure}

%\newpage
\begin{figure}[!b]
\centering
\caption{Coalescence of onions during a penetration scan 
at $\textrm{T} = 30^{\circ}\textrm{C}$. After 2 days (A) large connections form between
onions and (B) coalesced double onions with cores are observe. After 4 days (C)
big onion in the process of coalescence are observed. (D) Onion with domains 
viewed using partial cross polars. (E)
Closeup of the onion in (D).}
\label{hightemp2}
\end{figure}

%\newpage
\begin{figure}[!ht]
\centering
\caption{ Internal
defects in a diluted onion dispersion at $\textrm{T} = 30^{\circ}\textrm{C}$.  
(A) After $6$ hours of swelling 
(B) - (D) The defects in the onion disappear over 3 days when the temperature 
is changed to $\textrm{T} = 20^{\circ}\textrm{C}$.}
\label{quen}
\end{figure}
\vspace{8cm}

%\newpage
\begin{figure}[!hbt]
\centering
\caption{(A) Calculated density profile of an onion where $\phi$ is the bilayer density and $r$ is the radial position in an onion of radius $R$.    
(B) Pictorial representation of (A) where the dark regions
correspond to regions of low density and bright regions correspond to
regions of high density in the onion. (Figure courtesy of
H.~Diamant~\cite{diamant:cores}.)}
\label{haimscore}
\end{figure}

\end{document}